# Valley controlled propagation of pseudospin states in bulk metacrystal waveguides

Xiao-Dong Chen, Wei-Min Deng, Jin-Cheng Lu, and Jian-Wen Dong*

*School of Physics & State Key Laboratory of Optoelectronic Materials and Technologies, Sun Yat-sen University, Guangzhou 510275, China.*

*Corresponding author: dongjwen@mail.sysu.edu.cn

## ABSTRACT

Light manipulations such as spin-direction locking propagation, robust transport, quantum teleportation and reconfigurable electromagnetic pathways have been investigated at the boundaries of photonic systems. Recently by breaking Dirac cones in time-reversal invariant photonic crystals, valley-pseudospin coupled edge states have been employed to realize selective propagation of light. Here, without photonic boundaries, we realize the propagation of pseudospin states in three-dimensional bulk metacrystal waveguides by employing the ubiquitous valley degree of freedom. Valley-dependent pseudospin bands are achieved in three-dimensional metacrystal waveguides without Dirac cones. Reconfigurable photonic valley Hall effect is proposed after studying the variation of pseudospin states near K' and K valleys. Moreover, a prototype of photonic blocker is realized by cascading two inversion asymmetric metacrystal waveguides in which the pseudospin direction locking propagation exists. In addition, valley-dependent pseudospin bands are also discussed in a realistic metamaterials sample. These results show an alternative way towards molding the pseudospin flow in photonic systems.

## I. INTRODUCTION

By taking advantage of the various degrees of freedom such as frequency [1, 2], phase [3, 4], polarization [5] and momentum [6], the light flow control is of growing scientific and technological importance. Recently by considering the spin-orbit interaction in time-reversal-invariant photonic systems, the propagation of spin states at photonic boundaries has attracted much attention. Spin filtered effect and unidirectional transmission of spin states have been demonstrated in different photonic systems such as metasurfaces [7, 8], metallic slit [9], photonic crystal waveguides [10, 11], and chiral nanophotonic interfaces [12, 13]. In the past few years, topology has also been verified as a flexible degree of freedom (DoF) to mold the flow of light, and has provided great potential opportunities in photonics [14-18]. Protected by the bulk-edge correspondence [19], two counter-propagating gapless pseudospin-polarized edge states are found at the interfaces of two topologically distinct time-reversal systems [20-22]. By employing such exotic edge states, robust transport and even the reconfigurable detouring of pseudospin states have been demonstrated at photonic boundaries or domain walls [23-26]. It seems that a well-defined photonic boundary is necessary for the observation of photonic pseudospin propagation. Is it possible to control pseudospin flow in a bulk medium without photonic boundary?

On the other hand, valley, which labels the energy extrema of band structure at momentum space, has been employed to achieve a number of intriguing phenomena such as valley-selective Hall transport and circular dichroism in two-dimensional layered materials [27-30]. Regarding the similarity between electronic systems and classical systems, photonic and sonic counterparts of valley-Hall typed topological insulators have been investigated very recently [31-39]. Valley chirality locked beam splitting and topological transport of edge states were proposed and observed. Although most of the reported valley controlled behaviors are found in systems where Dirac cones are gapped, the two inequivalent but time-reversal K' and K valleys are ubiquitous in periodic triangular and honeycomb lattices, no matter whether Dirac cones present or not. It suggests that valley photonics and valley acoustics not only can be explored in systems possessing gapped Dirac cones, but also can be extended to general triangular and honeycomb systems.

In this work, we show valley controlled propagation of pseudospin states in 3D bulk metacrystal waveguides without Dirac cones. By breaking the inversion symmetry, we find valley-dependent pseudospin bands and the resultant pseudospin gap due to valley-pseudospin interaction. The variation of the pseudospin bands is shown in the plane of two constitutive parameters of metacrystal waveguides. Reconfigurable photonic valley Hall effect is then demonstrated by shifting the working frequency. Pseudospin direction locking propagation of pseudospin states is also illustrated, confirming that pseudospin-filtered feature can be achieved by using the valley DoF. Furthermore, a prototype of photonic blocker is proposed by cascading two metacrystal waveguides. In addition, valley-dependent pseudospin split bulk bands are also discussed in a realistic sample constructed by non-resonant and electromagnetic-dual metamaterials between two metal plates. These results show a way towards molding the flow of pseudospin states in photonic structures by using valley as an alternative binary DoF.

## II. PSEUDOSPIN STATES IN METACRYSTAL WAVEGUIDE

Figure 1(a) shows the schematic of a metacrystal waveguide consisting of one metacrystal and two parallel metal plates at $z = 0$ and $z = d_0$ (yellow plane). The unit cell of metacrystal (pink frame) has the hexagonal cross section with the size of $a_0$ and the height of $d_0$. Each cell is composed of three hexagonal rods which are indexed by 1 (green), 2 (cyan), and 3 (grey), respectively. For such a metacrystal waveguide, we have the following electromagnetic field solution for the first order guided modes,

$$\vec{E} = [e_x \sin(\frac{\pi}{d_0}z), e_y \sin(\frac{\pi}{d_0}z), -e_z \cos(\frac{\pi}{d_0}z)]^T$$

$$\vec{H} = [-h_x \cos(\frac{\pi}{d_0}z), -h_y \cos(\frac{\pi}{d_0}z), h_z \sin(\frac{\pi}{d_0}z)]^T \quad (1)$$

where $e_x$, $e_y$, $e_z$, $h_x$, $h_y$, and $h_z$ are functions of $(x, y)$ but $z$-direction invariant. With the definitions of $\vec{e} = (e_x, e_y, e_z)^T$ and $\vec{h} = (h_x, h_y, h_z)^T$, the Maxwell equations can be rewritten in a compact form, yielding,

$$\nabla \times \vec{e} = i\omega[\mu_0\mu_r\vec{h} + \xi_e\vec{e}], \quad \nabla \times \vec{h} = -i\omega[\varepsilon_0\varepsilon_r\vec{e} + \xi_e\vec{h}] \quad (2)$$

where $\xi_e$ is an effective bianisotropic tensor with $\xi_{e,xy} = \xi^*_{e,yx} = i\pi/\omega d_0$. Hence, the 3D metacrystal waveguide can be viewed as a 2D metacrystal with the pseudo-fields of ($\vec{e}$, $\vec{h}$) and a bianisotropic coefficient $\xi_e$ [23].

In order to construct decoupled pseudospin states, all hexagonal rods are assumed to be uniaxial (i.e., $\mu_{r,xx} = \mu_{r,yy}$) and electromagnetic-dual (i.e., $\boldsymbol{\varepsilon}_r = \rho \boldsymbol{\mu}_r$ with $\rho$ being a constant). The electromagnetic-dual symmetry guarantees the occurrence of photonic Kramer degeneracy [20, 40], resulting in the decomposition of the Maxwell equations into two decoupled pseudospins,

$$\nabla \times \begin{pmatrix} p_x^- \\ p_y^- \\ p_z^+ \end{pmatrix} = -i\frac{\omega}{c} \begin{pmatrix} \mu_{r,xx}\sqrt{\rho} & \xi_{e,xy} & \\ \xi_{e,yx} & \mu_{r,xx}\sqrt{\rho} & \\ & & -\mu_{r,zz}\sqrt{\rho} \end{pmatrix} \begin{pmatrix} p_x^- \\ p_y^- \\ p_z^+ \end{pmatrix}$$

$$\nabla \times \begin{pmatrix} p_x^+ \\ p_y^+ \\ p_z^- \end{pmatrix} = i\frac{\omega}{c} \begin{pmatrix} \mu_{r,xx}\sqrt{\rho} & \xi_{e,xy} & \\ \xi_{e,yx} & \mu_{r,xx}\sqrt{\rho} & \\ & & -\mu_{r,zz}\sqrt{\rho} \end{pmatrix} \begin{pmatrix} p_x^+ \\ p_y^+ \\ p_z^- \end{pmatrix} \quad (3)$$

where $\vec{p}^{\pm} = \sqrt{\rho\varepsilon_0}\vec{e} \pm \sqrt{\mu_0}\vec{h}$. Consequently, for the pseudospin-up states (↑) with nonzero ($p_x^-$, $p_y^-$, $p_z^+$), the in-plane components ($e_x$, $h_x$) or ($e_y$, $h_y$) are out-of-phase while the out-of-plane components ($e_z$, $h_z$) are in-phase [20, 41]. On the contrary, for the pseudospin-down states (↓) with nonzero ($p_x^+$, $p_y^+$, $p_z^-$), ($e_x$, $h_x$) or ($e_y$, $h_y$) are in-phase while ($e_z$, $h_z$) are out-of-phase. Although the pseudo-fields $\vec{e}$ and $\vec{h}$ may not be intuitive, they are closely related to the electromagnetic fields $\vec{E}$ and $\vec{H}$. For example, we write down $\vec{E}$ and $\vec{H}$ at $z = 3d_0/4$ from Eq. (1),

$$\vec{E} = [e_x \sin(3\pi/4), e_y \sin(3\pi/4), -e_z \cos(3\pi/4)]^T = [e_x, e_y, e_z]^T / \sqrt{2} \equiv \vec{e}/\sqrt{2}$$

$$\vec{H} = [-h_x \cos(3\pi/4), -h_y \cos(3\pi/4), h_z \sin(3\pi/4)]^T = [h_x, h_y, h_z]^T / \sqrt{2} \equiv \vec{h}/\sqrt{2} \quad (4)$$

It implies that the pseudo-fields ($\vec{e}$, $\vec{h}$) are linearly proportional to ($\vec{E}$, $\vec{H}$) at $z = 3d_0/4$ in metacrystal waveguide. As a result, the pseudospin of each state can be also defined by the phase relationship between ($E_x$, $H_x$), ($E_y$, $H_y$), or ($E_z$, $H_z$) at $z = 3d_0/4$.

Although the phase relationship between either in-plane or out-of-plane components can be used for pseudospin classification, we focus on the out-of-plane component throughout this paper. That is to say, the pseudospin-up state has in-phase ($E_z$, $H_z$) while the pseudospin-down state has out-of-phase ($E_z$, $H_z$).

In order to illustrate the pseudospin classification, we consider a conceptual 3D metacrystal waveguide. The left inset of Fig. 1(b) shows the unit cell of metacrystal which has the hexagonal cross section with the size of $a_0 = 30$ mm and the height of $d_0 = 48$ mm. In addition, three hexagonal rods are set with $\rho = 13$ and $\boldsymbol{\mu}_{r1} = $ diag$\{0.455, 0.455, 0.25\}$, $\boldsymbol{\mu}_{r2} = $ diag$\{0.67, 0.67, 0.25\}$, $\boldsymbol{\mu}_{r3} = $ diag$\{0.39, 0.39, 0.44\}$ [see more in Section III.E for the experimental design]. As a result, the electromagnetic-dual symmetry is fulfilled and the pseudospin states are well defined. Figure 1(b) shows the band structure of this inversion symmetry breaking metacrystal waveguide. As examples of pseudospin classification, we plot out the eigen-fields of four lowest photonic states at K point in Fig. 1(c), including the $E_z$, $H_z$, and their phase difference (PD, i.e., arg($E_z$)-arg($H_z$)) at $z = 3d_0/4$. Obviously, the 1$^{st}$ and 4$^{th}$ lowest states are pseudospin-up, as the $E_z$ and $H_z$ are in-phase and the resultant PD is 0. While the 2$^{nd}$ and 3$^{rd}$ lowest states are pseudospin-down as the PD is π. Thus for the band structure in Fig. 1(b), one can classify all states by marking pseudospin-up (pseudospin-down) states in blue (red) color. The photonic bands are doubly degenerate along the ΓM direction due to mirror symmetry protection. In contrast, the photonic bands with different pseudospin near K' and K valleys split in the frequency level. Around the frequency of 3 GHz, the pseudospin-down states are prohibited near

K' valley while the pseudospin-up states are prohibited near K valley. It leads to a valley dependent pseudospin gap, i.e., frequency range in which pseudospin-up (and equivalently pseudospin-down) states are allowed near one valley but prohibited near the other valley. When such pseudospin gap is frequency isolated, the frequency extrema makes valley an alternative DoF to manipulate the flow of pseudospin states in bulk metacrystal waveguides.

Note that according to the $C_3$-rotation eigenvalue of $E_z$ (or $H_z$) fields, the 1st and 2nd (3rd and 4th) lowest states at K point belong to $A$ ($E$) irreducible representation [42]. In this way, one can label the four lowest states with different group representations and pseudospin notations, e.g., $A^\uparrow$, $A^\downarrow$, $E^\uparrow$, and $E^\downarrow$, as shown in Fig. 1(c). It is distinct from the gapped Dirac cone cases where two $E^\uparrow$ and two $E^\downarrow$ states should be found. It implies that valley photonics or valley acoustics can be extended to general triangular and honeycomb systems beyond those possessing gapped Dirac cones.

## III. RESULTS AND DISCUSSIONS

### A. Phase diagram of pseudospin split bulk states

In this section, we discuss the phase diagram of pseudospin split bulk states near K' and K valleys, showing the evolution of pseudospin states. In a time-reversal invariant system, pseudospin-up and pseudospin-down states are always doubly degenerate when the system is inversion invariant. For example, when the rod 1 and rod 2 are of same constitutive parameters, [i.e., $\boldsymbol{\mu}_{r1} = \boldsymbol{\mu}_{r2} = \text{diag}\{0.67, 0.67, 0.25\}$, and see the left inset in Fig. 2(c)], the metacrystal waveguide is inversion symmetric. It results in the doubly degenerate frequency bands in the whole Brillouin zone [Fig. 2(c)]. Such pseudospin degeneracy can be lifted by breaking either time-reversal or inversion symmetry, and we consider the latter case which is more straightforward to realize. To break the inversion symmetry, we keep the constitutive parameters of rod 2 unchanged, but change the constitutive parameters of rod 1 (i.e., $\mu_{r1,xx}$ and $\mu_{r1,zz}$). After the inversion symmetry is broken, the pseudospin states near K' and K valleys will split due to the nonzero valley-pseudospin coupled interaction [40]. Pseudospin states near K' and K valleys evolve as functions of $\mu_{r1,xx}$ and $\mu_{r1,zz}$, and Figure 2(a) shows the phase diagram of pseudospin states near K valley. Note that those near K' valley can be predicted from the principle of time-reversal symmetry.

In Fig. 2(a), the solid black curve shows the accidental degeneracy between $A^\uparrow$ and $A^\downarrow$ states at K point. On the other hand, the dashed black curve shows the accidental degeneracy between $E^\uparrow$ and $E^\downarrow$ states at K point. These two curves divide the phase diagram into four domains which are indexed by Roman numbers from I to IV. In each domain, we plot the schematics of the second and third lowest bands near K valley. For example in domain I, both the second and third lowest bands are of pseudospin-down polarization [see two red bands in Fig. 2(a)]. Hence pseudospin-up states are prohibited near K valley and it leads to a pseudospin-up gap. One representative metacrystal waveguide with $\mu_{r1,xx} = 0.5$ and $\mu_{r1,zz} = 0.25$ in domain I is marked by the yellow dot in Fig. 2(a) and its band structure is shown in the top-left panel of Fig. 2(d). Numerical result proves once again that the second and third lowest bands near K valley are pseudospin-down polarized. The polarization of these two bands can be changed by altering $\mu_{r1,xx}$ and $\mu_{r1,zz}$, so as to reach other domains in Fig. 2(a). For example, we consider the metacrystal waveguide with $\mu_{r1,xx} = 0.74$ and $\mu_{r1,zz} = 0.38$ [marked by the orange dot in Fig. 2(a)]. When it goes from the yellow dot to the orange dot in the phase diagram, it passes through the solid black curve but not the dashed black curve. It implies that there is a mode exchange between $A^\uparrow$ and $A^\downarrow$ states, but no exchange between $E^\uparrow$ and $E^\downarrow$ states. Hence for metacrystal waveguides in domain II, the second lowest band changes to be pseudospin-up polarized while the third lowest band keeps as pseudospin-down polarized [blue band on the bottom while red band on the top in domain II in Fig. 2(a)]. This is in good agreement with the calculated band structure given in the top-right panel of Fig. 2(d). On the other hand, when the metacrystal waveguide goes from domain II to domain IV in the phase diagram, it passes through the dashed black curve. The mode exchange between $E^\uparrow$ and $E^\downarrow$ states happens and the third lowest band changes to be pseudospin-up polarized. As a result, both the second and third bands near K valley become pseudospin-up polarized [see two blue bands in domain IV in Fig. 2(a)]. This is confirmed by the band structure showing in the low-right panel of Fig. 2(d) for metacrystal waveguide with $\mu_{r1,xx} = 0.85$ and $\mu_{r1,zz} = 0.15$ (marked by the purple dot). Lastly in domain III, we consider the metacrystal waveguide with $\mu_{r1,xx} = 0.58$ and $\mu_{r1,zz} = 0.12$ (marked by the pink dot). Its band structure is plotted in the low-left panel of Fig. 2(d). The second lowest band near K valley is of pseudospin-down polarization, which is different to that of purple metacrystal waveguide in domain IV. This is because it experiences a mode exchange between $A^\uparrow$ and $A^\downarrow$ states when transforming from domain IV to domain III. Hence in the phase diagram in Fig. 2(a), four different combinations of the polarizations of the second and third bands near K valley can be found. It indicates a potential way to control pseudospin flow by manipulating the polarizations of pseudospin states and pseudospin gaps (e.g., reconfigurable photonic valley Hall effect presented in Sec. IIIB).

Note that the bandwidth of pseudospin gap can be enlarged by the accidental degeneracy between the 2$^{nd}$ and 3$^{rd}$ pseudospin states at K point (or equivalently K' point). As an example to achieve such accidental case, we keep $\mu_{r1,zz}$ = 0.25 but alter $\mu_{r1,xx}$. Figure 2(b) shows the frequency spectra of four pseudospin states at K point as a function of $\mu_{r1,xx}$. One can see that frequencies of these four K valley states increase with the decreasing of $\mu_{r1,xx}$. When $\mu_{r1,xx}$ = 0.67 at the cyan dot, the $A^\uparrow$ ($E^\uparrow$) state superposes to the $A^\downarrow$ ($E^\downarrow$) state due to the inversion invariance [Fig. 2(c)]. It is interesting to find that two pseudospin-down states (i.e., $A^\downarrow$ and $E^\downarrow$ states) are accidentally degenerate at $\mu_{r1,xx}$ = 0.455 (marked by the green dot). It leads to a pseudospin gap with a 12% gap-midgap ratio [Fig. 1(b)], enabling the broadband pseudospin flow control.

### B. Reconfigurable photonic valley Hall effect

One of the characteristic manifestations of valley controlled propagation of pseudospin states is the photonic valley Hall effect (PVHE) in which pseudospin states at different valleys can be separately routed. Employing opposite group velocities of pseudospin states in the second and third bands, reconfigurable PVHE can be achieved by shifting the working frequency. To see this, we consider the metacrystal waveguide in domain IV, e.g., that marked by the purple dot in the phase diagram [Figs. 2 and 3(a)]. As presented in the low-right panel in Fig. 2(d), the second and third bands near K valley are pseudospin-up polarized, while those near K' valley are of pseudospin-down polarization. Two pseudospin gaps, i.e., one ranging from 2.48 to 2.6 GHz and the other from 2.85 to 3 GHz, are found. As to determine the propagation directions of pseudospin states in these two gaps, equi-frequency contours should be considered. As examples, Figures 3(b) and 3(c) show the equi-frequency contours at the frequency of 2.48 GHz in the second band and the frequency of 2.92 GHz in the third band, respectively. These two contours are similar, but the pseudospin states on them propagate along opposite directions. For example, pseudospin-up states with $f$ = 2.48 GHz propagate along the ΓK direction [blue arrow in Fig. 3(b)], while they switch to propagate along the ΓK' direction when the frequency of 2.92 GHz is considered [blue arrow in Fig. 3(c)]. This is because the propagation direction of pseudospin state, i.e., the group velocity, is perpendicular to the contour and points in the direction of increasing frequency. For the second band, the direction of increasing frequency points towards the valley center, while it points away from the valley center for the third band. With these opposite group velocities, reconfigurable PVHE is expected [Figs. 3(d) and 3(e)]. An $E_x$ polarized source is launched into the metacrystal waveguide along the +$y$ direction [marked in yellow in Fig. 3(a)]. When the frequency of 2.48 GHz is considered, the pseudospin-up component from the source can be filtered out and routed up-leftwards along the ΓK direction, while the pseudospin-down component is transferred along the ΓK' direction. Such valley-dependent pseudospin-flow behavior is well demonstrated in Fig. 3(d) where the PD between $E_z$ and $H_z$ at $z = 3d_0/4$ is plotted. Obviously, the PD is stable around the value of 0 (cyan) at the end of ΓK propagating channel. Such in-phase feature indicates that nearly-pure pseudospin-up state propagates along the ΓK direction. On the contrary, the propagating waves along the ΓK' direction are pseudospin-down polarized as the PD is around π (red). On the other hand, as shown in Fig. 3(e), pseudospin-down (pseudospin-up) state will be obtained at the end of the ΓK (ΓK') propagating channel. This PD distribution is distinct to that shown in Fig. 3(d). With the comparison between Figs. 3(d) and 3(e), reconfigurable PVHE is achieved by shifting the operating frequency in the same metacrystal waveguide。

### C. Pseudospin direction locking propagation

The pseudospin direction locking propagation is also identified in Fig. 4. We consider the metacrystal waveguide in domain I, e.g., that marked by the green dot in the phase diagram [Figs. 1 and 4(a)]. When an $E_y$-polarized source is launched along the +$x$ direction, only pseudospin-down states propagating along the ΓK direction can be excited if the frequency of 2.9 GHz is considered. Figure 4(b) shows the $E_z$ fields of rightward propagating pseudospin state at $z = 3d_0/4$. The $E_z$ fields are parallel to $y$-axis at the right-exit. As only the pseudospin-down component of the incident source is filtered and transferred rightwards, the PD is stable around π [red in Fig. 4(c)]. In contrast, when the source is placed on the right, pseudospin-up state propagating leftwards along the ΓK' direction is excited [Fig. 4(d)]. This is verified by the result shown in Fig. 4(e) where the PD is stable around 0 (cyan) at the left-exit of metacrystal waveguide.

Note that such pseudospin direction locking propagation is unique in inversion asymmetric metacrystal waveguides. As a comparative case, we study the transmission in inversion symmetric metacrystal waveguide in Fig. 5. As shown in Fig. 5(a), we consider the metacrystal waveguide whose band structure has been presented in Fig. 2(c). As the inversion symmetry is kept, pseudospin-up and pseudospin-down states are doubly degenerate. Hence an $E_y$ incident source launching on the right will excite both left-ward pseudospin-up and pseudospin-down flow. These two pseudospin states interfere with each other when they propagate along the bulk crystal, resulting in the non-parallel output $E_z$ fields. This is demonstrated by the wavefront distortion at the left-exit in Fig. 5(b). As no pure pseudospin flow is obtained at the left exist, the PD distributions are messy and dependent on the y-positions.

### D. Prototype of photonic blocker

The valley-dependent pseudospin-split bulk band and the associated valley DoF open a route towards the discovery of novel states of light and fancy applications such as pseudospin-dependent light propagation, non-reciprocal transport of pseudospin states. In this section, we show the prototype of photonic blocker which is constructed by cascading two metacrystal waveguides [Fig. 6(a)]. The metacrystal waveguide locating on the right of the dashed yellow line is that presented in Figs. 1 and 3(a).

While the metacrystal waveguide on the left is obtained by inverting the right one by 180° along the *z* direction. When the incident source is placed on the right, it excites leftward pseudospin-up flow along the first metacrystal waveguide [see in Figs. 4(d) and 4(e)]. However, as the left metacrystal waveguide only support leftward pseudospin-down flow, the excited pseudospin-up flow in the right metacrystal waveguide will be reflected and refracted at the interface [see the bottom inset in Fig. 6(c)]. Hence, low transmittance will be observed at the left-exit and the photonic blocker can be realized. To test the performance of this proposed photonic blocker, we do the transmission simulation. As presented in Fig. 6(b), the excited light flow by right incident source is strongly reflected or refracted, and it leads to the enhanced $E_z$ fields at the right hand side. Nearly zero fields are observed at the left hand side of this blocker. We also calculate the transmittance of the photonic blocker and study the no-blocker case [i.e., Fig. 4(c)] for comparison. The transmittance of the photonic blocker case is two orders of magnitude lower than that of the non-blocker case [Fig. 6(c)].

**E. Experimental design of metacrystal waveguide**

In this section, we will present a concrete design for the proposed 3D metacrystal waveguide. As stated above, the 3D metacrystal waveguide consists of two parallel metal plates at *z* = 0 mm and *z* = 48 mm, and one sandwiched metacrystal with the height of 48 mm. As shown in Fig. 7(a), the designed metacrystal consists of six layers of metamaterials along the *z* direction. Each single-layer has the height of 8 mm, and it consists of one plexiglass plate on the bottom, one array of meta-atoms in the middle, and another plexiglass plate on the top [Fig. 7(b)]. The plexiglass plates have the height of 3 mm, and they are drilled with a honeycomb lattice (lattice constant of 30 mm) of through holes (diameter of 12 mm). As to put meta-atom array, blind holes with the height of 2.1 mm should be also drilled at the center of the pre-drilled honeycomb through holes. The geometries of these blind holes are depend on the meta-atoms putting on them. Between two plexiglass plates, meta-atoms with 'gyro' or 'star' geometries are put [Fig. 7(c)]. As shown in the top panel of Fig. 7(c), the gyro meta-atom consists of two concentric metallic cylinders with different diameters and heights, i.e., the fat-short cylinder with ($d_1$ = 26.8 mm, $h_1$ = 1 mm) and the thin-tall cylinder with ($d_2$ = 5.4 mm, $h_2$ = 6.2 mm). The star meta-atom is constructed by three same metallic blocks each rotated 60° with respect to one another. The size of metallic block is 18.7mm*4.6mm*6.2mm [bottom panel of Fig. 7(c)]. With these well-designed gyro and star meta-atoms, we construct the unit cell of meta-atom array by surrounding one star meta-atom with two gyro meta-atoms [outlined by the dashed black hexagon in Fig. 7(b)].

In short, as to construct the realistic 3D metacrystal waveguide, we first array the meta-atoms, and sandwich them between two plexiglass plates to form one single-layer of metamaterials, then stack six layers of metamaterials along the $z$ direction, lastly put two metal plates at $z$ = 0 and $z$ = 48 mm. Figure 7(d) shows the corresponding band structure of this inversion symmetric metacrystal waveguide. Protected by both the time-reversal and inversion symmetries, pseudospin states are doubly degenerate in the whole Brillouin zone. As to break the inversion symmetry, we reduce the diameter of short-fat cylinder ($d_1$) in the gyro meta-atoms locating in rod 1 region [labelled in green in Fig. 7(b)]. Frequency spectra of the four pseudospin states at K point as a function of $d_1$ are shown in Fig. 7(e). At $d_1 = 26.8$ mm [marked by the cyan dot], $A^\uparrow$ and $A^\downarrow$ ($E^\uparrow$ and $E^\downarrow$) states are frequency degenerate as the inversion symmetry is preserved. With the decreasing of $d_1$, frequencies of the four pseudospin states increase. Particularly, two pseudospin-down states are accidentally degenerate when $d_1 = 21.9$ mm (marked by the green dot). The bandwidth of pseudospin gap is enlarged. Hence, metacrystal waveguide with $d_1 = 21.9$ mm is chosen and its band structure is shown in Fig. 7(f). Similar to the band structure of metacrystal waveguide with effective parameters presented in Fig. 1(b), bands along the ΓM direction are nearly degenerate but pseudospin split bulk bands are found near K' and K valleys. Note that although pseudospin states at M point are not exactly degenerate, it would not affect the valley controlled propagation of pseudospin state as the related frequency region is far from the frequency range of interest. Besides, we also plot out the eigen-fields at $z = 3d_0/4$ of four pseudospin states at K point. Both the irreducible representation and pseudospin polarization of each pseudospin state are in good agreement with the theoretical results in Fig. 1(c). From the band structure and the eigen-fields in Fig. 7, we expect that the above-mentioned valley controlled behaviors, such as PVHE, pseudospin direction locking propagation, and photonic blocker, can be experimentally observed in this designed metacrystal waveguide around 3 GHz.

## IV. CONCLUSION

In conclusion, we discuss the valley controlled propagation of pseudospin states in 3D bulk metacrystal waveguides without Dirac cones. Valley-dependent pseudospin bands and the phase diagram of pseudospin states are obtained in metacrystal waveguides by breaking the inversion symmetry and changing the constitutive parameters. With the phase diagram, reconfigurable photonic valley Hall effect is achieved by shifting the working frequency. Pseudospin direction locking propagation is also realized by using valley as an alternative binary DoF. Employing the pseudospin filtered feature of inversion asymmetric metacrystal waveguides, we further demonstrate a prototype of photonic blocker. Lastly, a realistic metamaterials design of the proposed 3D metacrystal waveguides is discussed.

We would like to emphasize that valley-dependent electromagnetic wave behaviors can also be exploited by breaking the inversion symmetry in other photonic systems such as three-dimensional photonic crystals, silicon-based metamaterials, and gyrotropic medium. It will pave a way to not only fundamental physics that is difficult to observe in electronic systems, but also next generation of optical communication devices based on pseudospin-dependent light propagation, and non-reciprocal transport of pseudospin states.

## ACKNOWLEDGEMENTS

This work is supported by Natural Science Foundation of China (11522437),

Guangdong Natural Science Funds for Distinguished Young Scholar (S2013050015694), Guangdong special support program, and Fundamental Research Funds for the Central Universities (No.17lgpy19).

**FIGURES AND FIGURE CAPTIONS**

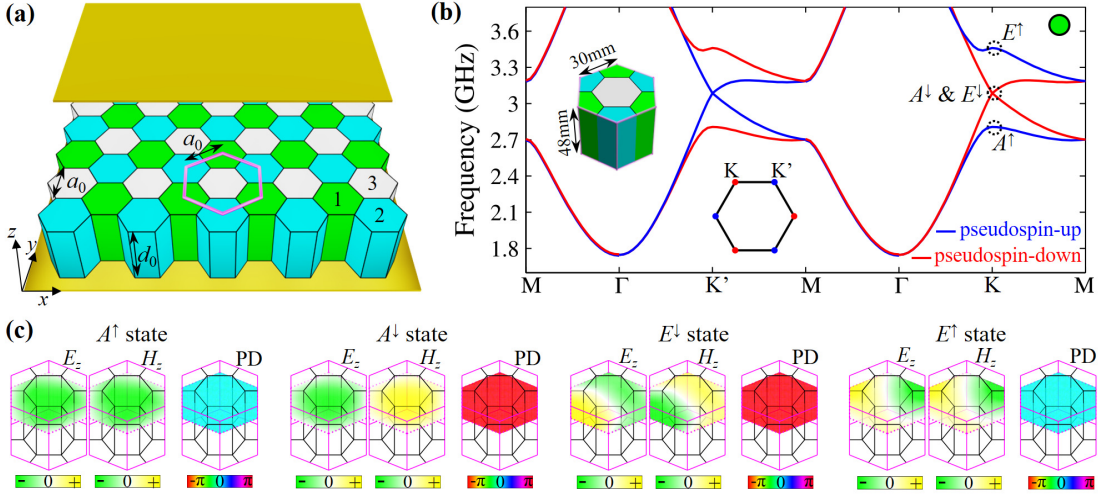

FIG. 1. (Color online) Valley-dependent pseudospin states in metacrystal waveguide. (a) Schematic of the inversion asymmetric metacrystal waveguide which is constructed by one electromagnetic-dual metacrystal and two parallel metal plates. The top plate is shifted to reveal the metacrystal inside. The unit cell of metacrystal has the height of $d_0$ along the $z$ direction and a hexagonal cross section with the size of $a_0$ in $xy$ plane (framed in pink). Each cell is composed of three hexagonal rods which are respectively indexed by 1 (green), 2 (cyan), and 3 (grey). (b) Valley-dependent pseudospin bands of metacrystal waveguide whose feature lengths are $a_0 = 30$ mm and $d_0 = 48$ mm (see the unit cell in the left inset). Three hexagonal rods are set with $\rho = 13$ and $\boldsymbol{\mu}_{r1} =$ diag$\{0.455, 0.455, 0.25\}$ [green rod], $\boldsymbol{\mu}_{r2} =$ diag$\{0.67, 0.67, 0.25\}$ [cyan rod], $\boldsymbol{\mu}_{r3} =$ diag$\{0.39, 0.39, 0.44\}$ [grey rod]. Pseudospin-up and pseudospin-down states are marked in blue and red, respectively. Around the frequency of 3 GHz, pseudospin-down states are prohibited near K' valley, resulting in a pseudospin-down gap. Similarly, pseudospin-up states are not allowed near K valley, and it leads to a pseudospin-up gap. The Brillouin zone with high symmetry K' and K points is shown in the middle inset. (c) Eigen-fields of the four lowest pseudospin states at K point. For each state, the ($E_z$, $H_z$) and phase difference (PD) between $E_z$ and $H_z$ at $z = 3d_0/4$ are shown. According to the $C_3$-rotation eigenvalues of $E_z$ (or $H_z$) fields, the 1st and 2nd (3rd and 4th) states belong to $A$ ($E$) irreducible representation. Inferring from the phase relation between $E_z$ and $H_z$ (or the resultant PD), the 1st and 4th states are pseudospin-up polarized while the 2nd and 3rd states are pseudospin-down polarized. As a result, these four states can be respectively labelled as $A^\uparrow$, $A^\downarrow$, $E^\uparrow$, and $E^\downarrow$ states according to the group representations and pseudospin notations [labelling upon the eigen-fields in (c) and outlined by dashed circles along the band structure in (b)]. Note that $A^\downarrow$ and $E^\downarrow$ states are accidentally degenerate at K point by carefully choosing $\mu_{r1,xx} = 0.455$ [see details in Fig. 2(b)]. This accidental degeneration is different to the structural degeneracy of Dirac point at which two $E$ states are considered.

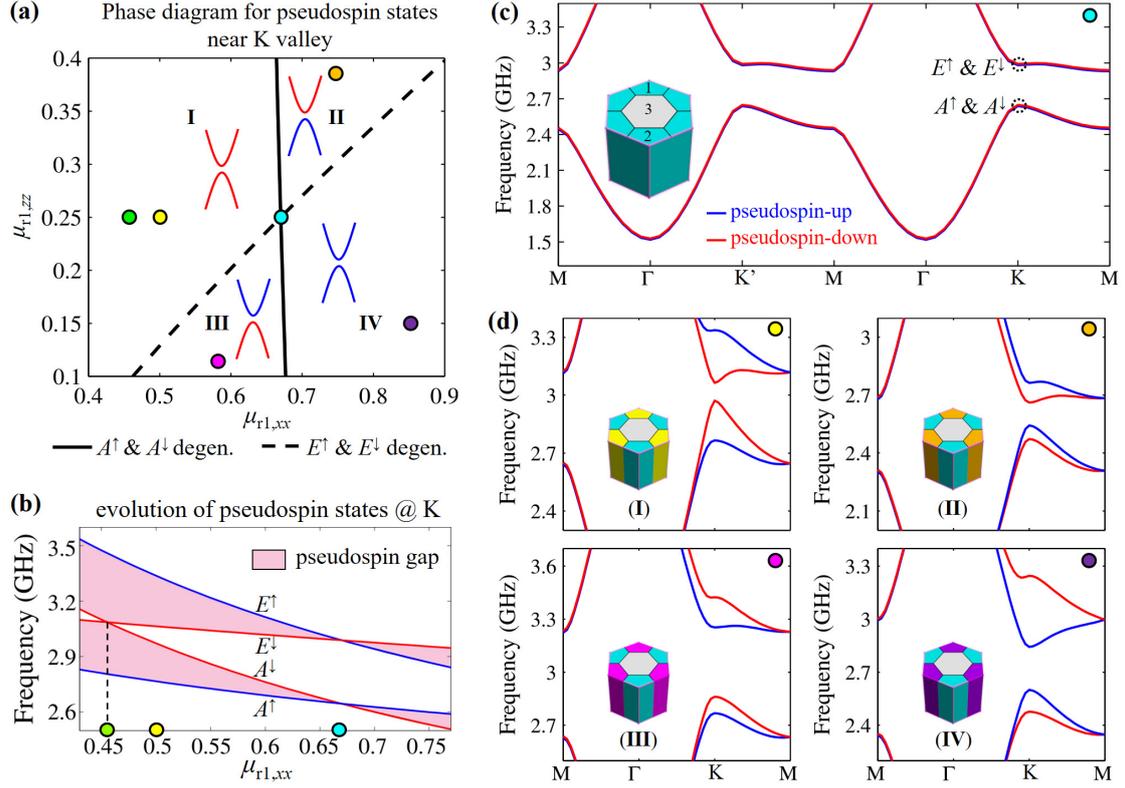

FIG. 2. (Color online) Variation of pseudospin states near K valley and band structures of representative metacrystal waveguides. (a) Phase diagram for pseudospin states near K valley by considering metacrystal waveguides with different $\mu_{r1,xx}$ and $\mu_{r1,zz}$. Here, only the phase diagram and schematics of band structures near K valley are shown, and those near K' valley are well predicted according to the time-reversal symmetry. The solid black (dashed black) curve shows the frequency accidental degeneracy between $A^\uparrow$ and $A^\downarrow$ ($E^\uparrow$ and $E^\downarrow$) states at K point. These two curves divide the phase diagram into four domains (indexed by Roman numbers from I to IV) which are characterized by different polarizations of the second and third lowest bands near K valley. (b) Frequency spectra of four pseudospin states at K point as a function of $\mu_{r1,xx}$, while $\mu_{r1,zz}$ is fixed at 0.25. By achieving the accidental degeneracy between the $A^\downarrow$ and $E^\downarrow$ states, the pseudospin gap bandwidth is maximized at $\mu_{r1,xx} = 0.455$. The green, yellow and cyan dots are in accordance with those in the phase diagram in (a). (c) Degenerate band structure of the inversion symmetric metacrystal waveguide whose unit cell is shown in the left inset. Three hexagonal rods within the unit cell are marked by indices of 1, 2, and 3, respectively. Rod 1 and rod 2 are set as $\boldsymbol{\mu_{r1}} = \boldsymbol{\mu_{r2}} = \mathrm{diag}\{0.67, 0.67, 0.25\}$ while the rod 3 is $\boldsymbol{\mu_{r3}} = \mathrm{diag}\{0.39, 0.39, 0.44\}$. (d) Band structures of four representative metacrystal waveguides in each domain, i.e., metacrystal waveguide with $\mu_{r1,xx} = 0.5$ and $\mu_{r1,zz} = 0.25$ (marked by the yellow dot in domain I), with $\mu_{r1,xx} = 0.74$ and $\mu_{r1,zz} = 0.38$ (marked by the orange dot in domain II), with $\mu_{r1,xx} = 0.58$ and $\mu_{r1,zz} = 0.12$ (marked by the pink dot in domain III), and with $\mu_{r1,xx} = 0.85$ and $\mu_{r1,zz} = 0.15$ (marked by the purple dot in domain IV).

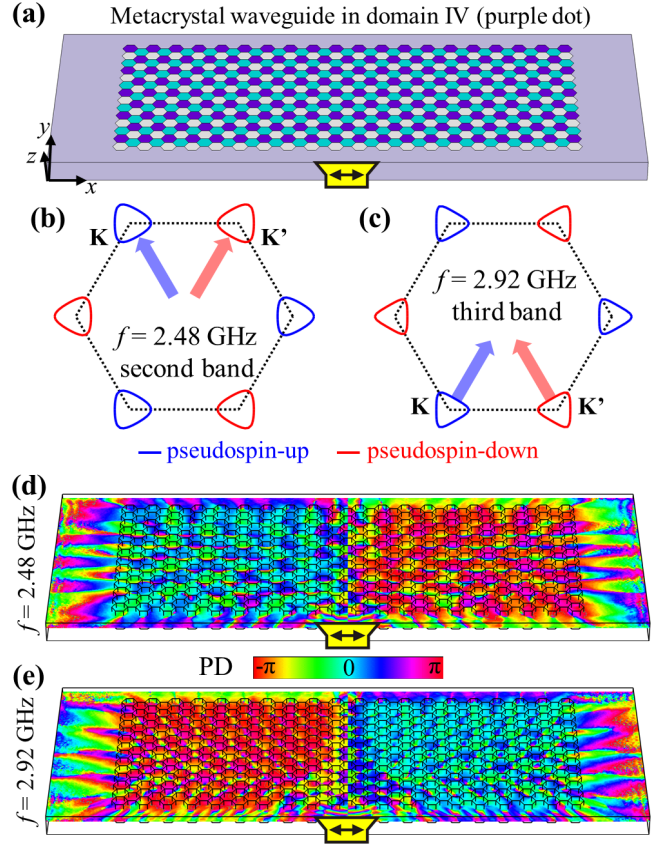

FIG. 3. (Color online) Reconfigurable photonic valley Hall effect in bulk metacrystal waveguide. (a) Schematic of the metacrystal waveguide in domain IV of Fig. 2(a). Outer light-purple region is the homogeneous dielectric medium with ε = 13 for guiding pseudospin states with matched impedance. An $E_x$ source is excited along the +y direction (marked in yellow). (b, c) Equi-frequency contour at the frequency of (b) 2.48 GHz and (c) 2.92 GHz. Pseudospin-up states at these two frequencies have similar contours but different increasing frequency directions, resulting in opposite group velocities (marked by blue arrows). Opposite group velocities are also found in pseudospin-down states (marked by red arrows). (d, e) The PD distributions when the incident source is operated at the frequency of (d) $f$ = 2.48 GHz, and (e) $f$ = 2.92 GHz. In (d), pseudospin-up states propagate along the ΓK direction while pseudospin-down states along the ΓK' direction. However, in (e), the pseudospin flow directions are reversed, and hence reconfigurable PVHE is confirmed.

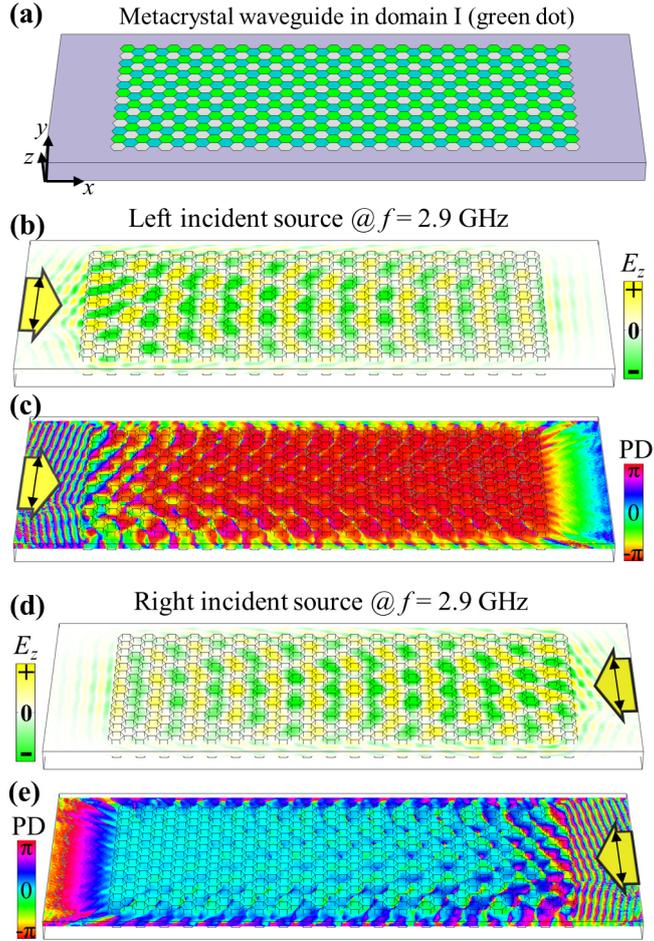

FIG. 4. (Color online) Pseudospin direction locking propagation in bulk metacrystal waveguide. (a) Schematic of the metacrystal waveguide which has been illustrated in Fig. 1 and marked by the green dot in the phase diagram in Fig. 2. An $E_y$ polarized source is launched along the $+x$ or $-x$ direction, with the operating frequency being $f = 2.9$ GHz. (b, c) When the source is launched on the left, only rightward pseudospin-down state propagating along the ΓK direction is excited, and the PD is around π (red color). (d, e) On the contrary, when the source is incident on the right, leftward propagating pseudospin-up state along the ΓK' direction is excited, and the PD is almost 0 (cyan color).

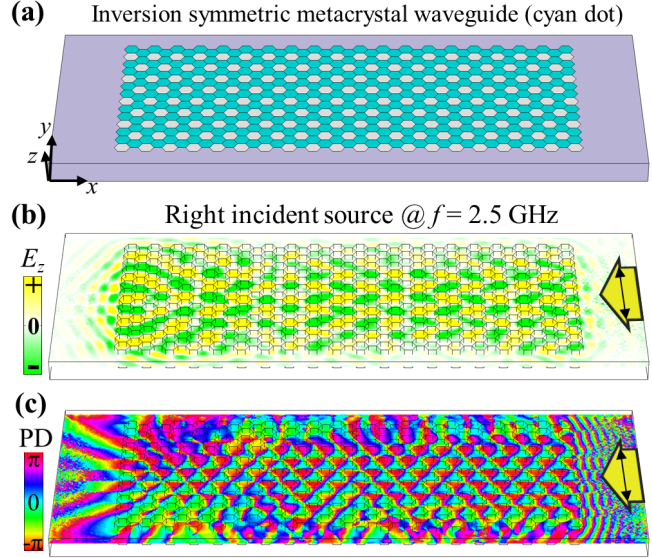

FIG. 5. (Color online) Absence of pseudospin direction locking propagation in the inversion symmetric metacrystal waveguide. (a) Schematic of the inversion symmetric metacrystal waveguide as that shown in Fig. 2(c). (b) Non-parallel output $E_z$ fields and (c) messy PD distributions at the left-exit when an $E_y$ polarized source is incident on the right, with the frequency of $f$ = 2.5 GHz. This is because both pseudospin-up and pseudospin-down states are excited, and they interfere with each other while propagating along the bulk crystal.

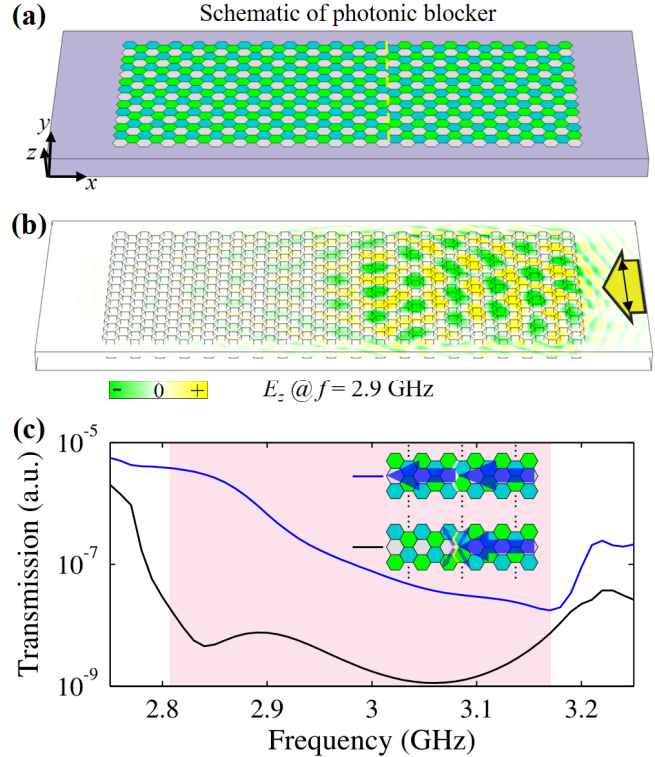

FIG. 6. (Color online) Prototype of photonic blocker. (a) Schematic of photonic blocker which is constructed by cascading two metacrystal waveguides together. (b) The $E_z$ fields for photonic blocker at the frequency of 2.9 GHz. Low transmittance is observed in the left-exit of the propagating channel. (c) Transmittance spectra for no-blocker (blue) and photonic blocker (black). For the no-blocker case, the leftward pseudospin-up flow meets no obstacles (top inset) while the excited pseudospin-up flow is reflected and refracted at the interface (bottom inset). Hence, the transmittance of the photonic blocker case is two orders of magnitude lower than that of the non-blocker case.

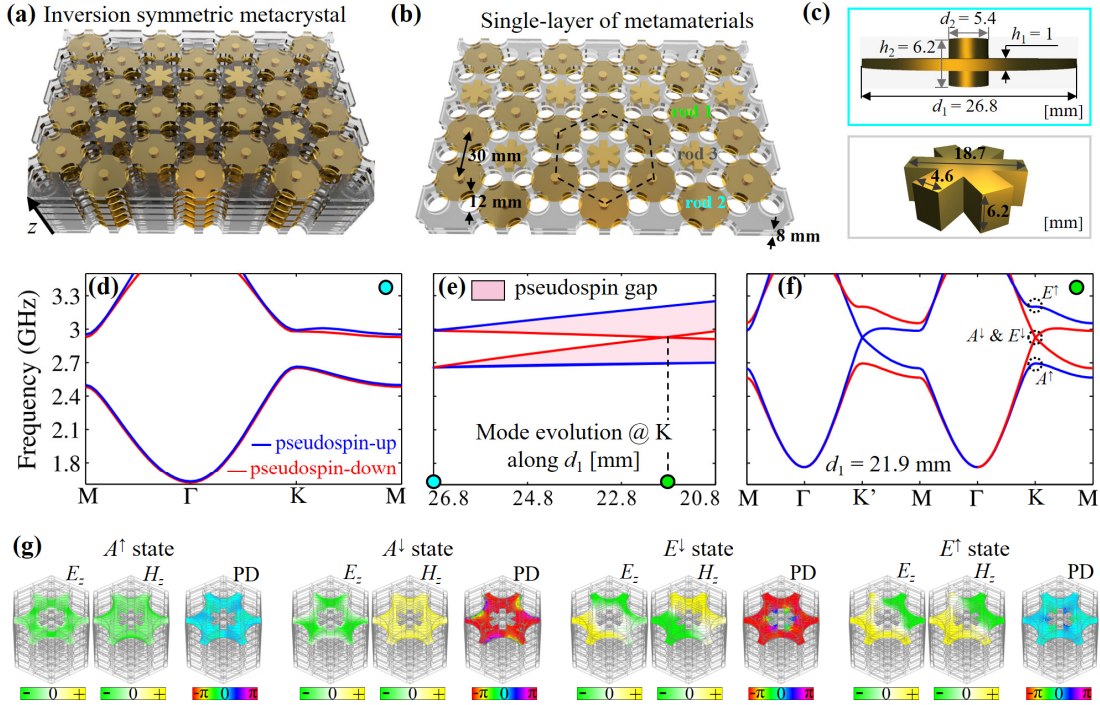

FIG. 7. (Color online) Experimental design and band structure of the realistic metacrystal waveguide. (a) Schematic of the inversion symmetric metacrystal consisting of six layers of metamaterials which are stacked along the $z$ direction. (b) Schematic of one single-layer of metamaterials with the height of 8 mm. It consists of two plexiglass plates with drilled holes and one array of meta-atoms. The unit cell of meta-atom array consists one star meta-atom surrounded by two gyro meta-atoms (dashed black hexagon). In accordance with the conceptual structure proposed in Figs. 1 and 2, the rod 1, rod 2, and rod 3 regions are labelled. (c) Schematics and structural parameters for the gyro (top panel) and star (bottom panel) meta-atoms. (d) Pseudospin degenerate band structure for inversion symmetric metacrystal waveguide. (e) Frequency spectra of four pseudospin states at K point as a function of the diameter of short-fat cylinder ($d_1$) in the gyro meta-atoms locating in rod 1 region. (f) Valley-dependent pseudospin split band structure for metacrystal waveguide with $d_1$ = 21.9 mm. All bands are marked in blue (pseudospin-up) or red (pseudospin-down) according to the pseudospin polarizations of eigen-states on them. Frequency isolated pseudospin split bulk bands appears near K' and K valleys. (g) Eigen-fields of four pseudospin states at K point, sharing the same irreducible representations and pseudospin notations as those in Fig. 1(c).